\pdfoutput=1

\documentclass[3p,times,twocolumn]{elsarticle}

\usepackage{ecrc}

\volume{00}

\firstpage{1}

\journalname{Nuclear Physics B Proceedings Supplement}

\runauth{}

\jid{nuphbp}

\jnltitlelogo{Nuclear Physics B Proceedings Supplement}

\usepackage{amssymb}
\usepackage[figuresright]{rotating}

\begin{document}

\begin{frontmatter}

\dochead{}

\title{A software package for stellar and solar inverse-Compton emission: Stellarics}

\author[label1]{Elena Orlando}
\author[label2]{Andrew Strong}
\address[label1]{W.W. Hansen Experimental Physics Laboratory, Kavli Institute for Particle Astrophysics and Cosmology, Stanford University, Stanford, CA, 94305, USA}
\address[label2]{Max-Planck-Institut fuer extraterrestrische Physik, Postfach 1312, 85741, Garching, Germany}

\begin{abstract}
We present  our software to compute gamma-ray emission from inverse-Compton scattering  by cosmic-ray leptons in  the heliosphere, as well as in the photospheres of stars. 
It includes a formulation of modulation in the heliosphere, but  can be used for any user-defined modulation model. 
Profiles and spectra  are output to FITS files in a variety of forms for convenient use. 
Also included are general-purpose inverse-Compton routines with other features like energy loss rates and emissivity for any user-defined target photon and lepton spectra.
The software is publicly available and it is under continuing development.
\end{abstract}


\end{frontmatter}

\section{Introduction}
\label{Introduction}
The importance of gamma-ray emission by inverse-Compton (IC) scattering of cosmic-ray
electrons and positrons on the photon field of individuals stars 
was first realized in 2006  \citep{Orlando2006, Orlando2007}. 
The same mechanism for emission coming from the Sun was found to be even more interesting.
In fact, the Sun was predicted
to be an extended source of gamma-ray emission,
produced by IC scattering of cosmic-ray
electrons on solar photons in \cite{Orlando2006}, \cite{Moskalenko} and  \cite{Orlando2007}. 
Following the discovery of solar emission from the quiet sun in EGRET
data \cite{Orlando2008}, now Fermi-LAT is so sensitive that even such weak emission can be
detected with high significance and studied in detail \cite{Abdo2011}. 
Hence, propagation of leptons in the inner heliosphere can be investigated, which is otherwise impossible.
Inverse-Compton emission has a broad distribution on the sky with maximum intensity in the direction of the Sun, but extending over
the entire sky at a low level. 
Hence, solar IC is also important as a background over the 
sky to be accounted for in studies of Galactic and extragalactic gamma-ray emission.

Calculations of the  Galactic IC
emission have usually assumed a smooth interstellar radiation field,
but in fact a large part of the Galactic luminosity
comes from the most luminous stars, which are
rare. Therefore we expect \cite{Orlando2006, Orlando2009} the interstellar radiation field, and hence the
inverse Compton emission, to be clumpy. 
Also the predicted IC  emission 
from Cygnus OB2 \cite{Orlando2007} can  be  modeled for example as 
in \cite{Orlando2009},
and contributes to the the emission from the Cygnus region observed by Fermi-LAT  \cite{Abdo_cygnus}.

It is important to have a reliable reference IC model available for general use.
We present here our C++ software to compute IC scattering from
the heliosphere, as well as the photospheres of stars. It includes a formulation of
modulation in the heliosphere, but can be used for any user-defined
modulation model. It outputs angular profiles and spectra  to
FITS files in table format in a variety of forms for convenient use.
The software is publicly available\footnote{http://sourceforge.net/projects/stellarics} and is
under continuing development.

\section{The software package}
\label{software}
The software computes inverse-Compton scattering from the heliosphere and the photosphere of individual stars.
The software uses general-purpose inverse-Compton routines with other features like energy loss rates and emissivity for any user-defined target photon and lepton spectra.
It is written in C++ and  is modular.
There are C++  classes for IC cross sections, emissivity spectra, stellar radiation fields, electron and positron spectra.
It can compute  both  isotropic and anisotropic IC scattering. 
A driver routine is provided and can be  adapted by the user as required.
New models of electrons, positrons and modulation can be easily added.
It has an optimized emissivity computation for spectral integrations and it can be run in parallel; the computation parameters allow a user-defined compromise between high resolution  and reasonable computation time.
It is very flexible since it contains user-defined parameters (energy ranges, integration steps etc.). 
Output is provided as FITS files (in table format), in various forms: angular profiles, spectra and differential and integrated flux. 
Output as $idl$ commands  is also provided.
 It has been tested with GNU and Intel compilers, and  optionally uses OpenMP for faster execution on multiprocessor machines.
A sample output dataset is provided with the package to  illustrate the format and check the installation.

\subsection{Main program}
\label{modulation}
The software package includes an example ($solar\_ic\_driver.cc$), which contains the parameters that can be edited by users depending on their requirements. 
This parameters to be defined are: energy range and grid factor for both cosmic-ray electrons and gamma rays, grid spacing (linear or logarithmic) of the angle from the Sun or the star, integration range and steps, parameters of the star (temperature, radius, distance), electron spectra models including cases with free parametrizations.

\subsection{Calculation}
\label{calculation}
The calculation of the IC emission integrated along the line of sight is contained in the class $SolarIC$. 
It uses the input parameters from $solar\_ic\_driver.cc$, the electron spectrum parametrization from $LeptonSpectrum$, the photon field of the star or Sun from $StarPhotonField$, as a black-body source, and the IC cross-sections based on the Klein-Nishina formula from $InverseCompton$. 
The option to use a logarithmic angular grid is useful to resolve the rapid variation near the solar disk while still covering the full angular range required.

\subsection{Electron spectrum}
\label{modulation}
The electron and positron spectra can be defined by the user by adding named models to the $LeptonSpectrum$ class. 
The model parameters are freely definable. 
Several sample models are provided, including those based on existing publications \citep{Orlando2008, Moskalenko, Abdo2011}. 
They will be regularly updated with new models.
Modulation is described with the force-field approximation, but this can be freely extended to more realistic models.

\subsection{Inverse Compton differential cross-section}
\label{modulation}
Cross-sections are calculated in class $InverseCompton$. 
Both isotropic and anisotropic Klein-Nishina cross-sections are included, as described in \cite{Moska2000, Orlando2008}. 
The Thompson cross-section is also included for a cross-check. 
The IC emissivity spectrum for a given lepton spectrum and radiation field is computed here, and this can be used for other applications such as interstellar emission.
An energy-loss computation for a given radiation field is also available here.

\begin{table*}
\begin{center}
\caption{Extensions names and contents of the output FITS file and corresponding  units. 
$I(E,\theta)$ is the differential intensity for gamma-ray energy $E$ at angle $\theta$ from the solar centre.}
----------------------------------------------------------------------------------------------------------------------------------------\\
\begin{tabular}{llccl}
Extension's Number and Name    &        Contents&        Row &    Column  & Units\\
                   &                &  Variable  &   Variable &      \\
\\
\\
1.     Differential\_intensity\_profile          &         $I( E, \theta)$      &   $E$     &       $\theta$   &       cm$^{-2}$ sr$^{-1}$ s$^{-1}$ MeV$^{-1}$\\
2. Energy\_integrated\_profile                    &        $I(>E, \theta)$     &      $E$       &     $\theta$    &    cm$^{-2}$ sr$^{-1}$ s$^{-1}$ \\
3.  Angle\_integrated\_profile                     &       $I( E,<\theta)$     &     $E$       &     $\theta$    &     cm$^{-2}$ s$^{-1}$ MeV$^{-1}$\\
4.  Energy\_and\_angle\_integrated\_profile   &             $I(>E,<\theta)$     &  $E$    &       $\theta$      &      cm$^{-2}$ s$^{-1}$\\
5.    Spectrum\_for\_angles                            &  $I( E,\theta)$     &      $\theta$   &     $E$        &     cm$^{-2}$ sr$^{-1}$ s$^{-1}$ MeV\\
6.    Spectrum\_times\_Esquared\_for\_angles   &            $I( E,\theta) \times E^{2}$  &     $\theta$ & $E$ &     cm$^{-2}$ sr$^{-1}$ s$^{-1}$ MeV\\
7.   Spectrum\_for\_integrated\_angles                 & $I( E,<\theta)$      &     $\theta$  &      $E$       &     cm$^{-2}$ s$^{-1}$ MeV$^{-1}$\\
8.  Spectrum\_times\_Esquared\_for\_integrated\_angles & $I( E,<\theta) \times E^{2}$   &    $\theta$ & $E$ &    cm$^{-2}$ s$^{-1}$ MeV\\
9.     Energies                               &         Energies&    $E$              &                   &            MeV\\
10.     Angles                                        &      Angles &  $\theta$       &                 &            degrees\\
\label{Table1}
\end{tabular}
----------------------------------------------------------------------------------------------------------------------------------------\\
\end{center}
\end{table*}

\section{Output}
\label{output}
The output is a FITS file with multiple extensions for convenient use e.g. directly with HEASARC's $fv$ tool, or  other plotting tools.
Extension names and contents of the output FITS file and respectively units are listed in Table \ref{Table1}.
 Alternatively the FITS file can be read by any program using $cfitsio$, or output as ASCII using $fv$. 
To plot a differential solar spectrum, as for example in Figure \ref{fig1}, run a solar model, choose extension 6 and plot the appropriate column against angle.
To plot solar  profiles of integral intensity choose extension 2, and plot the appropriate column against angle, as in Figure \ref{fig2}.

In addition, the program outputs an $idl$ program which generates  spectra and profiles directly: this option has been used to generate the plots in this article.

\section{Examples}
\label{examples}

\begin{figure}
\includegraphics[width=18pc, angle=0]{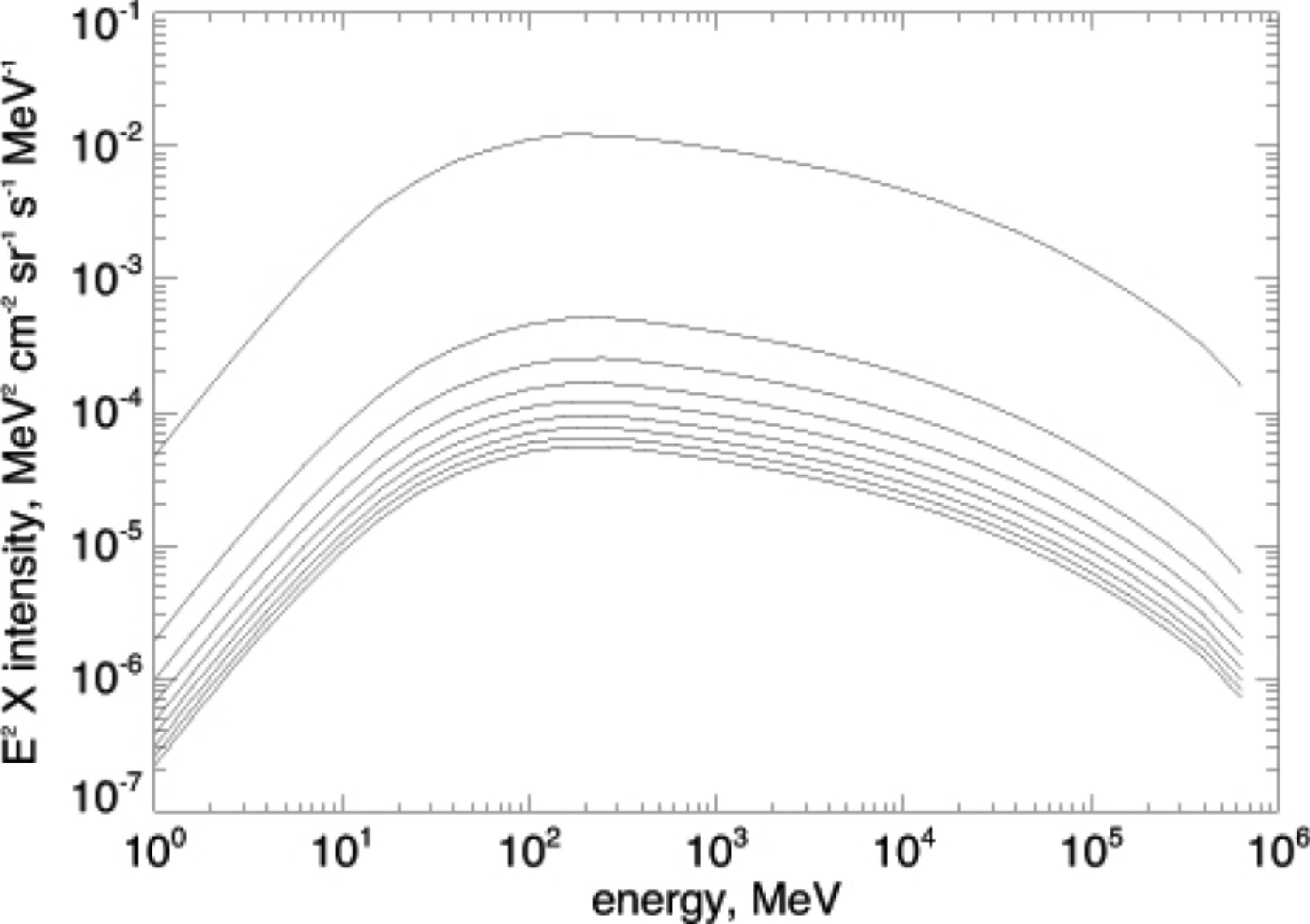}
\caption{Example of output: Intensity spectrum of solar IC at (from top) $0.3^o, 5^o, 10^o, 20^o,30^o,40^o$ from the Sun's center.  }
\label{fig1}
\end{figure}

\begin{figure}
\includegraphics[width=18pc, angle=0]{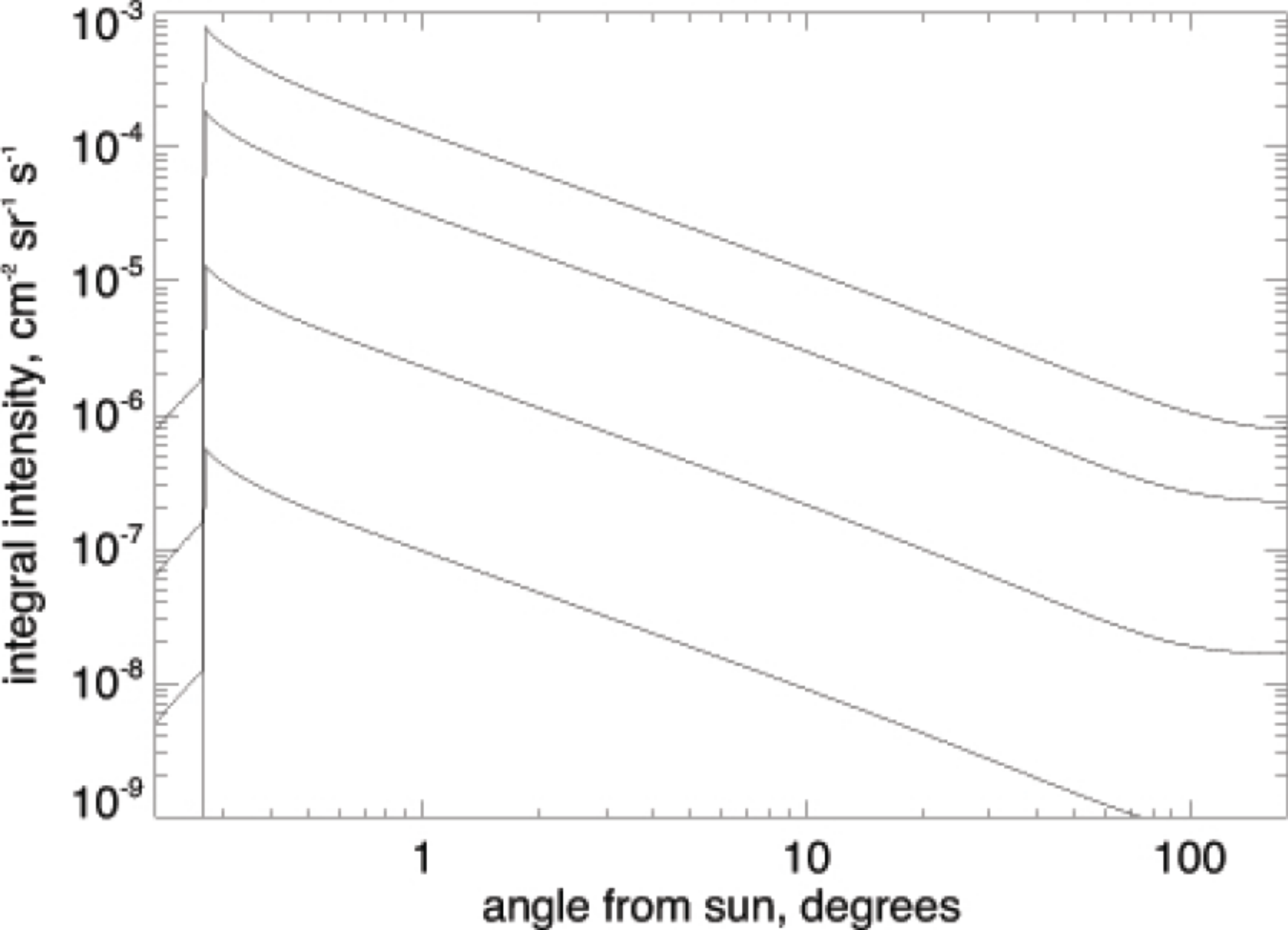}
\caption{Example of output: Angular profile of the IC emission as a function of the angular distance from the Sun, (from top) $>10$ MeV, $>100$ MeV, $>1$~GeV and $>10$ GeV. }
\label{fig2}
\end{figure}

Figs~\ref{fig1} and \ref{fig2} show the solar spectrum at various angles from the sun, and intensity profiles above various energies,  for a typical CR lepton spectrum. 
For the profiles a logarithmic angular grid was used to resolve the behaviour near the solar disk.

This plot illustrates the extension of the emission to the entire sky, and also demonstrates an interesting feature: due to the anisotropy of the solar radiation field the emission is almost zero towards the solar disk, the emission beyond the sun  being occulted. This effect should be visible in future high-quality Fermi-LAT data, and can help to distinguish the surface emission (cosmic-ray hadronic interactions) from the IC component.

\section{Stellar spectra}

Spectra for stars  have been presented in \citep{Orlando2006, Orlando2007}, and can be obtained with the present package by defining the relevant stellar parameters: luminosity, temperature and distance.

\section{Conclusions}
We have presented  our software for calculating the IC  emission from single stars and the Sun.
 The software is under continuing development, taking into account updated observations in gamma rays and cosmic rays. 
It is used as a standard  model for generating the solar IC emission  as a background  component and it will be especially useful for evaluating future data from Fermi-LAT.
\\

 Acknowledgements:
E.O. acknowledges support through NASA grants NNX12A073G.

\nocite{*}
\bibliographystyle{elsarticle-num}

\begin{thebibliography}{00}
\bibitem[Abdo et al.(2011)]{Abdo2011} Abdo, A.~A., Ackermann, 
M., Ajello, M., et al.\ 2011, ApJ, 734, 116 
\bibitem[Ackermann et al.(2011)]{Abdo_cygnus} Ackermann, M., 
Ajello, M., Allafort, A., et al.\ 2011, Science, 334, 1103 
\bibitem[Moskalenko 
\& Strong(2000)]{Moska2000} Moskalenko, I.~V., \& Strong, A.~W.\ 2000, ApJ, 528, 357 
\bibitem[Moskalenko et al.(2006)]{Moskalenko} Moskalenko, I.~V., 
Porter, T.~A., \& Digel, S.~W.\ 2006, ApJL, 652, L65 
\bibitem[Orlando 
\& Strong(2006)]{Orlando2006} Orlando, E., \& Strong, A.~W.\ 2006, arXiv:astro-ph/0607563
\bibitem[Orlando 
\& Strong(2007)]{Orlando2007} Orlando, E., \& Strong, A.~W.\ 2007, Ap\&SS, 309, 359 
\bibitem[Orlando 
\& Strong(2008)]{Orlando2008} Orlando, E., \& Strong, A.~W.\ 2008, A\&A, 480, 847 
\bibitem[Orlando 
\& Strong(2008)]{Orlando2009} Orlando, E., \& Strong, A.~W.\ 2008, International Cosmic Ray Conference, 2, 505 
\end{thebibliography}

\end{document}